\documentclass[prb, aps, onecolumn, showpacs]{revtex4}
\usepackage{amssymb}
\usepackage{amsmath}
\usepackage{braket}
\usepackage[pdftex]{graphicx}
\usepackage[english]{babel}
\usepackage[colorlinks]{hyperref} 
\begin{document}

\title{Effects of photon statistics in wave mixing on a single qubit}
\author{W. V. Pogosov$^{1,2}$, A. Yu. Dmitriev$^{3}$, O. V. Astafiev$^{4,3,5,6}$}
\affiliation{$^1$Dukhov Research Institute of Automatics (VNIIA), 127055 Moscow, Russia}
\affiliation{$^2$Institute for Theoretical and Applied Electrodynamics, Russian Academy of
Sciences, 125412 Moscow, Russia}
\affiliation{$^3$Laboratory of Artificial Quantum Systems, Moscow Institute of Physics and Technology, 141700 Dolgoprudny, Russia}
\affiliation{$^4$Skolkovo Institute of Science and Technology, 121205 Moscow, Russia}
\affiliation{$^5$Physics Department, Royal Holloway, University of London, Egham, Surrey TW20 0EX, United Kingdom}
\affiliation{$^6$National Physical Laboratory, Teddington, TW11 0LW, United Kingdom}

\begin{abstract}
We theoretically consider wave mixing under the irradiation of a
single qubit by two photon fields. The first signal is a classical
monochromatic drive, while the second one is a nonclassical light.
Particularly, we address two examples of a nonclassical light: (i) a
broadband squeezed light and (ii) a periodically excited quantum
superposition of Fock states with 0 and 1 photons. The mixing of
classical and nonclassical photon fields gives rise to side peaks
due to the elastic multiphoton scattering. We show that side peaks
structure is distinct from the situation when two classical fields
are mixed. The most striking feature is that some peaks are absent.
The analysis of peak amplitudes can be used to probe photon
statistics in the nonclassical mode.
\end{abstract}

\author{}
\maketitle
\date{\today }

\section{Introduction}

Wave mixing is a well known phenomenon in the domain of nonlinear
optics that has various applications \cite{0,00,000}. This effect
manifests itself in a generation of waves with new frequencies as a
result of interaction between incoming two or three frequency waves,
which conserves the total energy of photons. Wave mixing occurs in
nonlinear medium characterized by nonzero second-order or
higher-order susceptibilities \cite{00}.

Recent progress in microfabrication methods and quantum fields
control resulted in the possibility to realize nonlinear effects on
the level of a single artificial quantum system. Progress in this
direction is of importance in the context of quantum information
processing. One of the promising platforms for the construction of
quantum devices is superconducting quantum circuits. Particularly,
superconducting systems offer regimes which are not accessible for
natural atoms and give rise to various unusual quantum optics
phenomena both in on-chip and open-space configurations, see, e.g.,
Refs. \cite{Wallraff,1,2,3,4,5,Siddiqi,7,DCE1,DCE2,Dima}. An example
of such a phenomena is a wave mixing on a single artificial atom
that was demonstrated experimentally in the series of articles
\cite{Dmitriev2017,Decrinis2018,Dmitriev}. The atom plays a role of
a nonlinear element providing interaction between microwaves. In
Ref. \cite{Dmitriev} wave mixing of continuous coherent waves on a
superconducting flux qubit coupled to the coplanar waveguide was
demonstrated and existence of narrow side peaks of different orders
in nonlinearity was observed, which have been attributed to elastic
multiphoton scattering. Although both the experimental and
theoretical results of Ref. \cite{Dmitriev} were obtained for
coherent waves only, it was suggested that amplitudes of side peaks,
in general, should be sensitive to photon statistics of incident
waves and this feature can be used to probe their statistical
properties. This could be realized by mixing classical and
nonclassical drivings on atom that should allow for the
reconstruction of information on
 quantum statistics in the nonclassical mode \cite{Dmitriev}.
 Note that four-wave mixing of two coupled light modes was
 theoretically proposed for the quantum non-demolition measurement of
 the photon number in a selected mode performed by destructive measurement
 of photons in another coupled mode  \cite{Milburn1983counting,walls1985analysis}.

Here, we theoretically consider wave mixing in the case of
nonclassical photon field. We address a dynamics of single qubit
irradiated simultaneously by the coherent wave and nonclassical
light. We consider two examples of nonclassical field that is
produced either by degenerate parametric
amplifier\cite{Gardiner1,Siddiqi} or by a single-photon
source\cite{singlephoton1,singlephoton2,singlephoton3}. We indeed
find that peaks structure is not identical to the case of wave
mixing of two continuous coherent waves -- for example, some peaks
turn out to be absent. For the case of single-photon source,  we get
the three-peaked spectrum which is similar to what was observed for
the case of classical driving trains of pulses with relative time
delay \cite{Dmitriev2017}. For the squeezed vacuum in one mode and a
classical drive in another mode, we get only peaks containing even
number of photons from squeezed mode, while other peaks are absent.
We conclude that the peak amplitudes can be used to probe the
statistical properties of incident waves.

The paper is organized as follows.  In Section II we consider the
wave mixing under the irradiation by two coherent waves along the
ideas of Ref. \cite{Dmitriev}. In Section III, we analyze similar
equations of motion for qubit degrees of freedom under the
irradiation by coherent wave and broadband squeezed light. In
Section IV, we consider wave mixing under the irradiation by the
coherent wave and a periodically excited superposition of Fock
states with 0 and 1 photons. We conclude in Section V.

\section{Wave mixing under the irradiation by two coherent waves}

\begin{figure}[h]\center
\includegraphics[width=.5\linewidth]{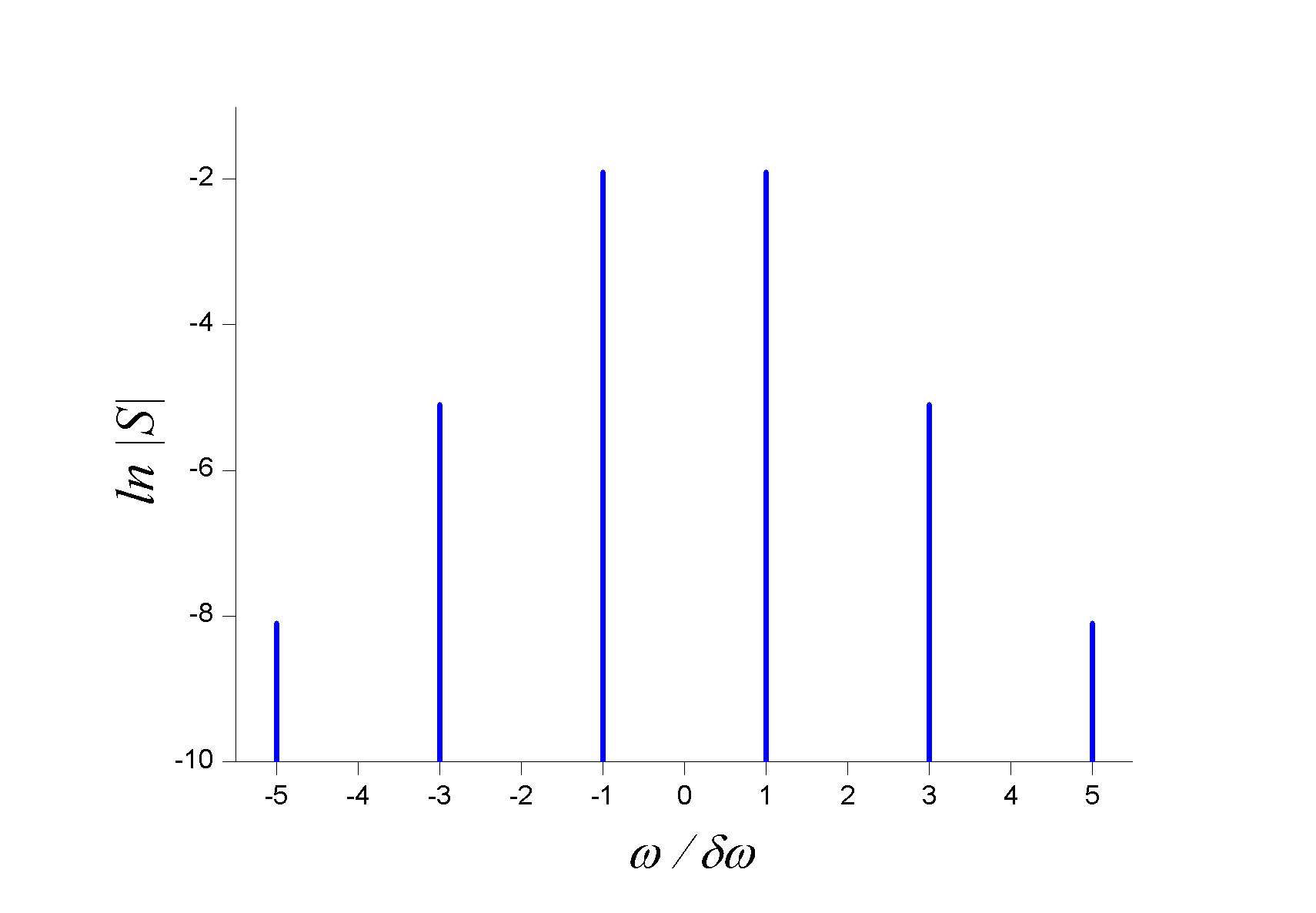}
\caption{Spectral components of $\langle \sigma_- \rangle$ in  the
case of qubit irradiation by two coherent waves (see in the text).}
\label{spectra}
\end{figure}

Let us reproduce main theoretical results of Ref. \cite{Dmitriev}.
We consider the dynamics of the qubit coupled to the transmission
line under the classical drive with two frequencies $\omega_{1}$ and
$\omega_{2}$ close to the qubit transition frequency $\omega_{01}$,
amplitudes of drives being $\Omega_1$ and $\Omega_2$, respectively.
The relaxation of the atom $\Gamma$ is radiative due to the photon
emission into the waveguide and the difference between $\omega_{1}$
and $\omega_{2}$ is much smaller than $\Gamma$,
$|\omega_{1}-\omega_{2}|\ll \Gamma$.

We switch to the rotating frame characterized by the frequency
$\omega_d = (\omega_{1}+\omega_{2})/2$ and introduce notations
$\delta \omega = \omega_{1} - \omega_d = \omega_d - \omega_{2}$.
Maxwell-Bloch  equations in this frame and under the rotating wave
approximation read as
\begin{equation}
\frac{d \langle \sigma_- \rangle}{dt} =  \langle \sigma_- \rangle
\left( -i \Delta \omega -\gamma \right)
 -\frac{i\Omega_1}{2} e^{-i \delta \omega t} \langle \sigma_z \rangle -\frac{i\Omega_2}{2} e^{i \delta \omega t} \langle \sigma_z \rangle, \label{MB1}
\end{equation}
\begin{equation}
\frac{d \langle \sigma_z \rangle}{dt} = -\Gamma (\langle
\sigma_z\rangle+1) + i \Omega_1 \left(\langle \sigma_+ \rangle e^{- i \delta \omega t}
- \langle \sigma_- \rangle e^{i \delta \omega t}\right) + i \Omega_2 \left(\langle \sigma_+ \rangle e^{ i \delta \omega t}
- \langle \sigma_- \rangle e^{ -i \delta \omega t}
 \right), \label{MB2}
\end{equation}
where $\Delta \omega = \omega_{01} - \omega_d$,  $\Gamma$ is the
radiative decay rate due to the coupling to the waveguide, while
$\gamma$ is a decoherence rate, which also depends on pure dephasing
rate $\Gamma_{\varphi}$: $\gamma=\Gamma/2+\Gamma_{\varphi}$.

It is straightforward to find a stationary  solution taking into
account that $\delta \omega t$ is a slowly varying phase on the
timescale of $\Gamma^{-1}$. This solution can be represented as
\begin{equation}
\langle \sigma_z \rangle = - \left(1+\frac{\gamma}{\Gamma}
\frac{\Omega_1^2+\Omega_2^2+\Omega_1 \Omega_2 (e^{-2i \delta \omega
t}+e^{2i \delta \omega t})}{\left(\Delta
\omega\right)^2+\gamma^2}\right)^{-1}, \label{MBsteadysigmaz}
\end{equation}
\begin{equation}
\langle \sigma_- \rangle = \frac{1}{2}  \frac{\Omega_1 e^{-i \delta
\omega t}+ \Omega_2 e^{i \delta \omega t}}{\Delta \omega - i \gamma}
\left(1+\frac{\gamma}{\Gamma} \frac{\Omega_1^2+\Omega_2^2+\Omega_1
\Omega_2 (e^{-2i \delta \omega t}+e^{2i \delta \omega
t})}{\left(\Delta \omega\right)^2+\gamma^2}\right)^{-1}.
\label{MBsteady}
\end{equation}
The amplitude of the elastically scattered wave is $-i\Gamma \langle
\sigma_- \rangle /\mu $, where $\mu$ is the qubit dipole moment
\cite{Dmitriev,Zagoskin}. It is clear from this result that
amplitudes of spectral components of the emitted power are nonzero
for all frequencies divisible by $\delta \omega$ (in the rotating
frame).  Particularly, Eq. (\ref{MBsteady}) can be rewritten as
\cite{Dmitriev}
\begin{equation}
\langle \sigma_- \rangle = \frac{\Omega_1 e^{-i \delta \omega t}+
\Omega_2 e^{i \delta \omega t}}{\Lambda} \tan \vartheta
\sum_{p=-\infty}^{+\infty} (-\tan (\vartheta/2))^{|p|} e^{i 2p
\delta \omega t}, \label{MBsteady1}
\end{equation}
where
\begin{equation}
\Lambda = \frac{4 \gamma \Omega_1 \Omega_2}{\Gamma (\Delta \omega +i \gamma)}, \label{MBsteady2}
\end{equation}
\begin{equation}
\vartheta = \arcsin \frac{2 \gamma \Omega_1 \Omega_2}{\Gamma \left(
(\Delta \omega)^2 + \gamma^2 \right) + \gamma
(\Omega_1^2+\Omega_2^2)}. \label{MBsteady3}
\end{equation}
Spectral components of $\langle \sigma_- \rangle$, defined through
$S(\omega) = \lim_{t\rightarrow \infty}
\frac{1}{t}\int_{-t/2}^{t/2}\langle \sigma_- \rangle \exp(-i\omega
t)dt$, are illustrated in Fig. \ref{spectra} at
$\Omega_1=\Omega_2=0.15 \Gamma$, $\Delta \omega =0$, $\Gamma =
2\gamma$. There was observed a quantitatively good agreement between
the theory and experimental results for the case of a
superconducting flux qubit irradiated by two coherent fields
\cite{Dmitriev}.

The wave mixing can be understood in terms of multiphoton elastic
scattering involving frequencies of photons from the coherent waves
\cite{00}, which correspond to arrows in Fig. \ref{spectra} --
lengths of arrows provide frequencies, while arrow directions show
either an absorption (up) or emission (down). The absorption of two
photons with frequencies $\omega_{1}$ and emission of a single
photon with $\omega_{2}$ produces the frequency
$2\omega_{1}-\omega_{2}=\omega_{d} + 3 \delta \omega$, since the
process is elastic and accompanied by the energy conservation. In
the same way, the absorption of two photons from the
$\omega_{2}$-mode and emission of a single from $\omega_{1}$-mode
gives rise to the peak at $2\omega_{2}-\omega_{1}=\omega_{d} - 3
\delta \omega$. These two processes correspond to the four-wave
mixing. Similarly, higher-order processes involving $2l+1$ photons
are possible that result in spectral peaks at frequencies
$(l+1)\omega_{1}-l \omega_{2}$ and $(l+1)\omega_{2}-l \omega_{1}$,
$l$ being an integer number. According to the idea of Ref.
\cite{Dmitriev}, the intensities of sidebands can be used to extract
information about photon statistics of incident waves.

Note that the wave mixing phenomenon is robust against energy
dissipation into degrees of freedom different from photon modes. In
this case, the amplitude of the elastically scattered wave is
determined by purely radiative relaxation rate, while qubit's
dynamics is described by Maxwell-Bloch equations with full $\Gamma$
and $\gamma$, which incorporate losses.

For optics in visible range, the single artificial atom is to be
replaced with a cloud of identical natural atoms to achieve a strong
coupling with propagating field. In this system there is a strong
resonant absorption, so the only experimentally available
configuration of bichromatic classical drive implies that
$\delta\omega \gg \Gamma$. For this case, the solution for elastic
and inelastic spectrum was analytically and numerically elaborated
in several works
\cite{ruyten1992elastic,Agarwal1991spectrum,freedhoff1990resfluor}.
Particularly, it was predicted that the elastic side peaks do appear
at combination frequencies $\omega_{\pm(2l+1)} = \omega_d \pm
(2l+1)\delta\omega$ and with intensities  proportional to
$J_{0}^2(2\Omega/\delta\omega)J_{2l+1}^2(2\Omega/\delta\omega)$,
where $\Omega_1=\Omega_2=\Omega$, but no experiments demonstrating
this dependence are known (here $J_l$ is $l$-th Bessel function of
the first kind). Here we consider the opposite case of small
$\delta\omega$ which is specifically appropriate for superconducting
qubits as frequency of single microwave tone is controlled with
great precision.

\section{Wave mixing under qubit irradiation by a coherent wave and squeezed light}

In this section we address the effect of the simultaneous
irradiation of the qubit by the classical coherent drive with
frequency $\omega_{1}$ and squeezed light. As the squeezed vacuum is
significantly non-classical and has non-trivial photon statistics
\cite{breichenbach1997}, the mixing of classical and squeezed
signals will result in a side components which are different from
ones for classical drives \cite{Dmitriev}. Thereby, wave mixing will
allow to investigate photon statistics in the nonclassical mode.

A paradigmatic example of a source of nonclassical light is
degenerate parametric amplifier described by the Hamiltonian of the
driven cavity \cite{Gardiner1}
\begin{eqnarray}
H_s =\omega_{2}a_{0}^{\dagger}a_{0} +  \frac{i}{2}\left(a_{0}^{\dagger 2} \epsilon_s
e^{-2i\omega_{2}t} - a_{0}^2 \epsilon_s^{*}
e^{2i\omega_{2}t}\right),\label{Hsqueezed}
\end{eqnarray}
where $a_{0}$ is the destruction operator for the internal cavity
mode with frequency $\omega_{2}$,  while a classical pump frequency
is also $\omega_{2}$. The output field of degenerate parametric
amplifier is a finite-bandwidth squeezed light. In the squeezed
white noise limit correlations functions of the output field can be
represented as \cite{Gardiner1}
\begin{equation}
\langle a_{out}^{\dagger} (t) a_{out} (t') \rangle = N e^{i \omega_{2} (t-t')} \delta (t-t'), \\
\label{sqcorrN}
\end{equation}
\begin{equation}
\langle a_{out} (t) a_{out} (t') \rangle = M e^{-i \omega_{2} (t+t')} \delta (t-t'),
\label{sqcorrM}
\end{equation}
where $M$ is a measure of light squeezing. In general, $|M|^2 \leq
N(N+1)$,  while $|M|^2 = N(N+1)$ corresponds to the pure squeezed
state.

The output field from the parametric amplifier is treated as an
input field for the qubit \cite{Gardiner1}. The interaction between
the qubit and the light is described by a usual electric-dipole
approximation: $\sum_{\omega}g_{k}(a_{k}^{\dagger} \sigma_- + a_{k}
\sigma_+)$. The equations for motion for the mean values $\langle
\sigma_- \rangle$ and $\langle \sigma_z \rangle$ in the case of a
qubit interacting with the output field from the degenerate
parametric amplifier having central frequency close to $\omega_{01}$
are generally known from literature
\cite{Gardiner1,Gardiner11,Gardiner2,Zoller}. They take a simple
form \cite{Gardiner1} in the white-noise limit, when the bandwidth
of the squeezed light significantly exceeds $\gamma$. We include
into consideration additional classical drive at another frequency
$\omega_{1}$ that is also close to $\omega_{01}$, as described by
the Hamiltonian $f_q(t) \sigma_x$, where
$f_q(t)=-\Omega_1(e^{i\omega_{1}t}+e^{-i\omega_{1}t})/2$. The
equations of motion in the white-noise limit read as (see, e.g., Eq.
(10.3.2) of Ref. \cite{Gardiner1})
\begin{equation}
\frac{d \langle \sigma_- \rangle}{dt} =  \langle \sigma_- \rangle
\left( -i \omega_{01} -\gamma(1+2N) \right) -\frac{i\Omega_1}{2}
e^{-i \omega_{1} t} \langle \sigma_z \rangle - 2\gamma M
e^{-2i\omega_{2}t} \langle \sigma_+ \rangle, \label{sigmaminussq}
\end{equation}
\begin{equation}
\frac{d \langle \sigma_z \rangle}{dt} = -\Gamma (\langle
\sigma_z\rangle+1) - 2N \Gamma \langle \sigma_z \rangle +  i\Omega_1
\left(\langle \sigma_+ \rangle e^{-i \omega_{1} t} - \langle
\sigma_- \rangle e^{i \omega_{1} t} \right). \label{sigmazsq}
\end{equation}
Note that the last term in the right-hand side  of Eq.
(\ref{sigmaminussq}) describes a process of absorption of a photon
pair accompanied by the qubit excitation. We again switch to the
rotating frame characterized by the frequency $\omega_d =
(\omega_{1}+\omega_{2})/2$ as in the case of two coherent fields.
The stationary solution in the rotating wave approximation is
\begin{equation}
\langle \sigma_z \rangle = - \frac{1}{1+2N}+ \frac{\gamma
\Omega_1^2}{\Gamma (1+2N)} \frac{1+ \frac{M}{2N+1} e^{4 i \delta
\omega t}+ \frac{M^*}{2N+1} e^{- 4 i \delta \omega t}}{(\Delta
\omega)^2 +\gamma^2\left[(2N+1)^2 -4|M|^2\right]+ \frac{\gamma
\Omega_1^2}{\Gamma}\left(1+ \frac{M}{2N+1} e^{4 i \delta \omega t}+
\frac{M^{*}}{2N+1} e^{- 4 i \delta \omega t}\right)},
\label{sigmaminussqsteady}
\end{equation}
\begin{equation}
\langle \sigma_- \rangle = \frac{\Omega_1}{2}\frac{(i \gamma +
\frac{\Delta \omega}{2N+1})e^{-i \delta \omega t}  + i \gamma
\frac{2M}{2N+1} e^{3 i \delta \omega t}}{(\Delta \omega)^2
+\gamma^2\left[(2N+1)^2 -4|M|^2\right]+ \frac{\gamma
\Omega_1^2}{\Gamma}\left(1+ \frac{M}{2N+1} e^{4 i \delta \omega t}+
\frac{M^{*}}{2N+1} e^{- 4 i \delta \omega t}\right)}.
\label{sigmaminussqsteady}
\end{equation}

We see from Eq. (\ref{sigmaminussqsteady}) that

(i) peaks structure in the spectrum is not identical to the similar
structure in the case of two coherent fields,

(ii) nonzero squeezing $M$ together with classical drive produces side peaks,

(iii) without classical drive, no peak appears under the irradiation by only a squeezed light.

\begin{figure}[h]\center
\includegraphics[width=.5\linewidth]{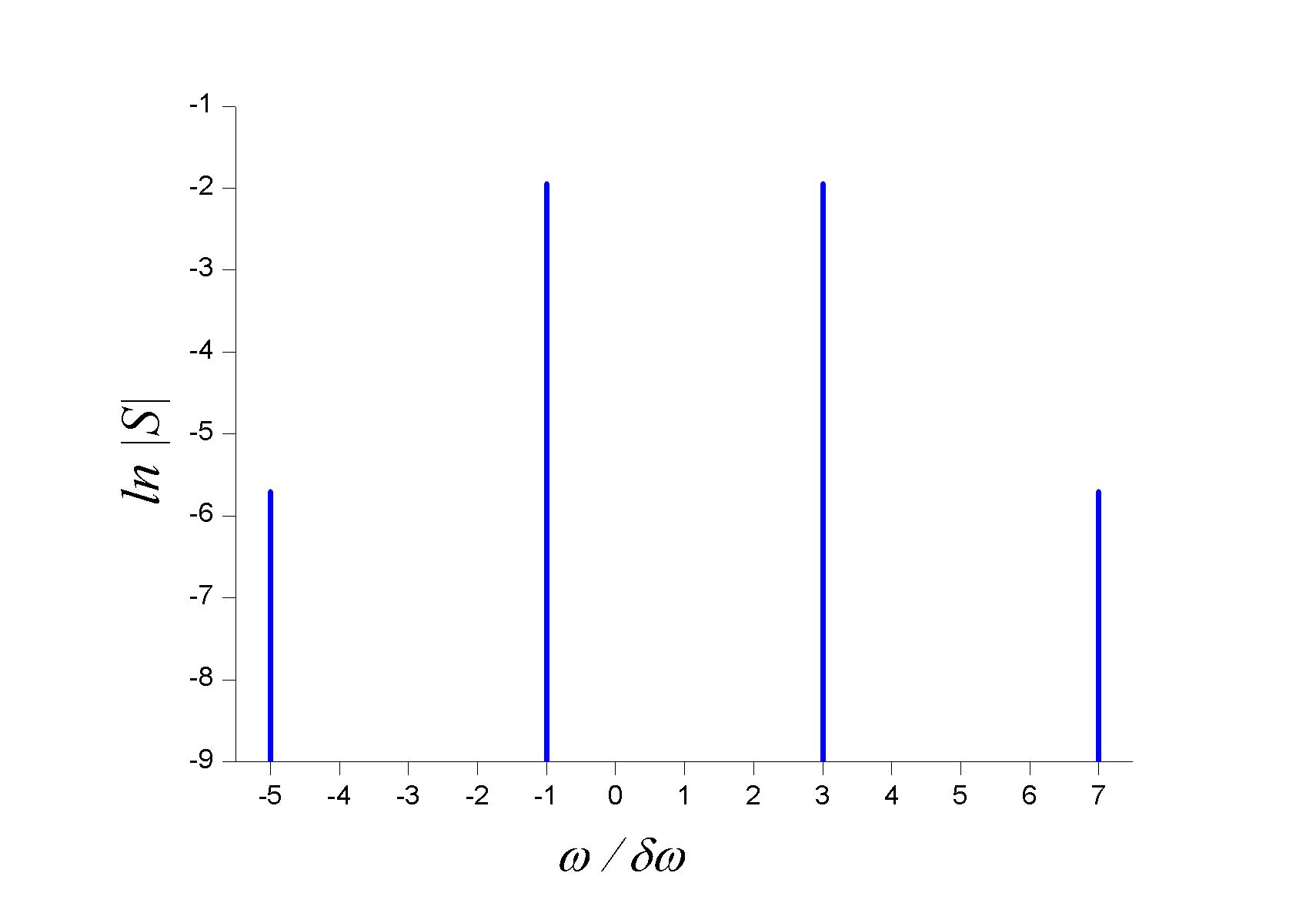}
\caption{Spectral components of  $\langle \sigma_- \rangle$ in the
case of qubit irradiation by a coherent wave together with the pure
squeezed light light (see in the text).} \label{spectra2}
\end{figure}

Spectral components $S(\omega)$ of $\langle \sigma_- \rangle$ are
shown in Fig. \ref{spectra2} at $\Omega_1 = 0.15 \Gamma$, $\Delta
\omega =0$, $\Gamma = 2\gamma$ and for the pure squeezed state with
$2|M|/(2N+1) \simeq 1$, $M$ being real. Compared to Fig
\ref{spectra}, the spectrum is shifted and some peaks are absent.

The obtained results can be qualitatively explained as follows. In
absence of a coherent drive the photon field at the qubit is just a
broadband output field from the degenerate parametric amplifier
which contains correlated photon pairs \cite{Gardiner1} each pair
having total energy $2 \omega_{2}$. This means that there is no
resonant frequency for a single photon since the radiated field is
broadband, but there is such a frequency for each correlated photon
pair. Therefore no peak appears in the spectrum of $\langle \sigma_-
\rangle$ without an additional coherent field. For the same reason
there is no peak at $\delta \omega$ also in presence of this field.

\begin{figure}[h]\center
\includegraphics[width=.6\linewidth]{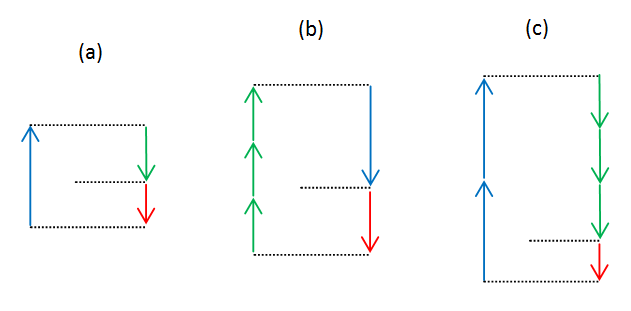}
\caption{Schematic images of multiphoton processes resulting in different side peaks in the emission spectra: at $3\delta \omega$ (a), $-5\delta \omega$ (b), $7\delta \omega$ (c). Blue arrows show absorption and emission of correlated photon pairs with total energy $2 \omega_{2}$. Green arrows correspond to single photons with energies $\omega_{1}$. Red arrows indicate photon emission, which is responsible for the side peaks.} \label{scheme}
\end{figure}

The first side peak appears at $3\delta \omega$ and it  corresponds
to the multiphoton process when a photon \emph{couple} with total
frequency $2 \omega_{2}$ is absorbed and a single photon with the
frequency $\omega_{1}$ is emitted, giving rise to the output photon
with $2 \omega_{2}-\omega_{1} = \omega_{d} - 3\delta \omega$. This
process is illustrated in Fig. \ref{scheme}(a). The dominant
contribution to the amplitude is proportional to both $\Omega_1 /
\Gamma$ and $M$, since $M$ provides a number of correlated photon
pairs in the incident nonclassical light. The peak at $-5\delta
\omega$ appears as a result of the absorption of three photons of
frequency $\omega_{1}$ and the emission of a photon couple having a
total frequency $2\omega_{2}$; this mechanism produces output
photons with frequency $3\omega_{1} - 2\omega_{2} =
\omega_{d}+5\delta \omega$, see Fig. \ref{scheme}(b). The amplitude
is proportional to the product of $M^{*}$ and $(\Omega_1 /
\Gamma)^3$. The peak at $7\delta \omega$ appears as a result of the
absorption of two pairs with $2\omega_{2}$ and the emission of three
photons with $\omega_{1}$; the amplitude is therefore proportional
to the product of $M^2$ and $(\Omega_1 / \Gamma)^3$, since two
correlated photon pairs are involved, see Fig. \ref{scheme}(c). And
so on.

In the weak driving regime,  $\Omega_1\ll \Gamma$, and at the
resonance, $\Delta \omega=0$, Eq. (\ref{sigmaminussqsteady}) can be
represented as
\begin{equation}
\langle \sigma_- \rangle \approx \frac{i f \Gamma}{\Omega_1}\left(f
e^{-i \delta \omega t} + f m e^{3 i \delta \omega t} - f^3 m^* e^{-5
i\delta \omega t} - f^3 m^2 e^{7i \delta \omega t} + \ldots
\right),\label{sigmaminussqsteady1}
\end{equation}
where we, for the simplicity, assumed that pure dephasing is
negligible, so that $\gamma = 2\Gamma$,
\begin{equation}
f = \frac{\Omega_1}{\sqrt{2\Gamma \gamma \left[(2N+1)^2 -4|M|^2\right]}},
\end{equation}
\begin{equation}
m = \frac{2M}{2N+1}.
\end{equation}
We see that apart of the general prefactor, $\langle \sigma_-
\rangle$ in the stationary  state is a sum of contributions, which
correspond to different multiphoton processes, each contribution
being proportional to $f$ in a power given by the number of photons
in the coherent wave participating in this process, as well as to
the squeezing characteristics $m$, which depends on the total number
of correlated pairs in the nonclassical wave, in a power given by
the number of correlated photon pairs also participating in a given
process. The side peaks at $\delta \omega (1+4l)$, where $l$ is an
arbitrary integer number, are absent, since there is no multiphoton
process that can produce these peaks. The obtained results also
evidence that the squeezing parameter can be reconstructed from the
analysis of side peaks amplitudes in the emission spectra - for
example, a direct comparison of two largest peak amplitudes at  $-
\delta \omega $ and $ 3 \delta \omega $ directly gives $m$.

\section{Wave mixing under qubit irradiation by a coherent wave and quantum superposition of vacuum and one photon}

In this Section we consider another example of wave mixing, when
nonclassical light is represented by periodically generated
superpositions of Fock states with 0 and 1 photons. We assume  that
the additional qubit serves as a emitter and creates mentioned
superpositions in the semi-infinite waveguide due to strong coupling
with the continuum of modes. This source for quantum superpositions
of vacuum and one photon can be engineered, for example, on the
basis of ideas of Refs. \cite{singlephoton1,singlephoton2}, where
tunable single-photon sources constructed from artificial
superconducting atoms were demonstrated. The emitter is periodically
excited by a strong external drive, which brings it to
quantum-mechanical superposition of lowest energy state
$\ket{\downarrow}$ and excited state $ \ket{\uparrow}$ with fixed
weights. The relaxation of the excited state is radiative due to the
single photon emission into the line. Let us denote a tunable
probability for the photon to be emitted after the excitation pulse
as $\nu$. The excitation pulse is assumed to be much shorter than
the emitter relaxation characteristic time $1/ \gamma_e$. Hereafter
indices $e$ are referred to the emitter. The time interval between
two excitation pulses $T$ is much larger than $1/ \gamma_e$. Using
Bloch sphere representation, the emitter state at $t=T n$, $n$ being
an integer number, can be expressed as
\begin{equation}
\langle \sigma_{-}^{e}(T n) \rangle  = \frac{\sin \theta}{2} e^{-i \omega_{01}^{e}Tn}, \label{emitter1}
\end{equation}
\begin{equation}
\langle
\sigma_{z}^{e} (T n) \rangle= \cos \theta, \label{emitter2}
\end{equation}
where $\theta$ is a polar angle, $\cos \theta = 2\nu -1$.
Equivalently,  the same state can be represented as $\sqrt{1-\nu}
\ket{\downarrow} + e^{-i \omega_{01}^{e}Tn}
\sqrt{\nu}\ket{\uparrow}$. The presence of the phase factor $e^{-i
\omega_{01}^{e}Tn}$ implies that the emitter is excited by the Rabi
pulse with the frequency $\omega_{01}^{e}$ coinciding with the
emitter transition frequency. The excitation is assumed not to alter
quantum state of the second qubit, which is responsible for the wave
mixing, that can be achieved in experiments by using different
methods. The emitter relaxation creates the superposition of 0 and 1
photons in the waveguide. These photon states then are mixed with
the continuous classical monochromatic drive of frequency
$\omega_{1}$ and amplitude $\Omega_1$ when they together irradiate
the second qubit characterized by the dissipation rate $\gamma \sim
\gamma_e$. Since the nonclassical signal at the second qubit's
position at any time instance contains no more than a single photon
($T \gg 1/\gamma$), side peaks structure must be distinct from the
case of wave mixing under two coherent drives, because higher orders
mixing processes cannot take place.

The presence of the qubit-emitter can be described by interaction
term\cite{Gardiner2}, which has a form $\sqrt{\gamma \gamma_e}
(\sigma_+ \sigma_-^{e}+\sigma_- \sigma_+^{e})$. It corresponds to
the interaction of two qubits via photon field treated in Markov
approximation and can be derived using, e.g., chain of equations of
motion in Heisenberg picture. Since much less attention has been
paid in literature for such a problem of qubit dynamics under the
irradiation from the quantum emitter, we include a microscopic
derivation of the equations of motion to Appendix A. This derivation
is based on Heisenberg equations of motion. The modification of the
Maxwell-Bloch equations now takes the form
\begin{equation}
\frac{d \langle \sigma_- \rangle}{dt} =  \langle \sigma_- \rangle
\left( -i \omega_{01} -\gamma \right)
  -\frac{i\Omega_1}{2} e^{-i \omega_{1} t} \langle \sigma_z \rangle +
    \sqrt{\gamma \gamma_e} \langle  \sigma_z \sigma_-^{e}\rangle, \label{MB3}
\end{equation}
\begin{equation}
\frac{d \langle \sigma_z \rangle}{dt} = -\Gamma (\langle
\sigma_z\rangle+1)  + i \Omega_1 \left(\langle \sigma_+ \rangle e^{-
i \omega_{1} t} - \langle \sigma_- \rangle e^{ i \omega_{1}
t}\right) + 2 \sqrt{\gamma \gamma_e} (\langle \sigma_- \sigma_+^{e}
\rangle + \langle \sigma_+ \sigma_-^{e}\rangle).
  \label{MB4}
\end{equation}
The right-hand sides of both equations  contain correlators $\langle
\sigma_z \sigma_-^{e}\rangle$, $\langle \sigma_- \sigma_+^{e}
\rangle$, $\langle \sigma_+ \sigma_-^{e}\rangle$, which cannot be
factorized due to the fact that we consider an ultraquantum limit
and this fact makes the situation distinct from the case of qubit
irradiation by classical signals. We treat these correlators as
follows. We consider first a steady state of the qubit under the
irradiation of only a classical drive and at $t=Tn$:
\begin{equation}
\langle \sigma_z (T n) \rangle = - \left(1+\frac{\gamma}{\Gamma} \frac{\Omega_1^2}{\left(\Delta \omega\right)^2+\gamma^2}\right)^{-1}. \label{detector1}
\end{equation}
\begin{equation}
\langle \sigma_- (T n) \rangle = \frac{1}{2} \frac{\Omega_1 e^{- i \omega_{1} Tn}}{\Delta \omega - i \gamma} \left(1+\frac{\gamma}{\Gamma} \frac{\Omega_1^2}{\left(\Delta \omega\right)^2+\gamma^2}\right)^{-1}. \label{detector2}
\end{equation}
Note that these two equations can be obtained from Eqs.
(\ref{MBsteadysigmaz}) and (\ref{MBsteady}) by assuming that the
amplitude of one of the classical signals is zero, $\Omega_2=0$. Now
we obtain from Eqs. (\ref{emitter1}), (\ref{emitter2}),
(\ref{detector1}), (\ref{detector2})
\begin{equation}
\langle  \sigma_z \sigma_-^{e} (Tn)\rangle = \langle  \sigma_z (Tn)
\rangle \langle \sigma_-^{e} (Tn)\rangle = -
\left(1+\frac{\gamma}{\Gamma} \frac{\Omega_1^2}{\left(\Delta
\omega\right)^2+\gamma^2}\right)^{-1} \frac{\sin \theta}{2} e^{-i
\omega_{01}^{e}Tn}. \label{correl1}
\end{equation}
\begin{equation}
\langle \sigma_- \sigma_+^{e} (T n) \rangle = \langle \sigma_- (T n)
\rangle  \langle \sigma_+^{e} (T n) \rangle = \frac{1}{2}
\frac{\Omega_1 }{\Delta \omega - i \gamma}
\left(1+\frac{\gamma}{\Gamma} \frac{\Omega_1^2}{\left(\Delta
\omega\right)^2+\gamma^2}\right)^{-1} \frac{\sin \theta}{2}
e^{i(\omega_{01}^{e} -\omega_{1} )Tn}. \label{correl2}
\end{equation}

The equations of motion for these correlators at $t \in (Tn,Tn+T)$ read as
\begin{equation}
\frac{d \langle  \sigma_z \sigma_-^{e} \rangle}{dt} =  \langle
\sigma_z \sigma_-^{e} \rangle \left(- i \omega_{01}^{e} - \Gamma -
\gamma_e \right), \label{EM1}
\end{equation}
\begin{equation}
\frac{d \langle \sigma_- \sigma_+^{e} \rangle }{dt} = \langle \sigma_- \sigma_+^{e} \rangle
\left( i(\omega_{01}^{e} -\omega_{01}) - \gamma - \gamma_e \right).
  \label{EM2}
\end{equation}
From these two equations we obtain
\begin{equation}
\langle  \sigma_z \sigma_-^{e} (t)\rangle =  - |\langle  \sigma_z \sigma_-^{e} (Tn)\rangle|
e^{-i \omega_{01}^{e}Tn}
e^{- i \omega_{01}^{e} (t-Tn)}
e^{-\left( \Gamma + \gamma_e \right) (t-Tn)}, \label{corr1}
\end{equation}
\begin{equation}
\langle \sigma_- \sigma_+^{e} (t)\rangle = |\langle \sigma_- \sigma_+^{e} (Tn)\rangle |
e^{i(\omega_{01}^{e} -\omega_{1} )Tn}
e^{i(\omega_{01}^{e} -\omega_{01})(t-Tn)}
e^{-(\gamma + \gamma_e)(t-Tn)}.
  \label{corr2}
\end{equation}

Note that, in principle, the dynamics of  correlators $\langle
\sigma_z \sigma_-^{e}\rangle$, $\langle \sigma_- \sigma_+^{e}
\rangle$, $\langle \sigma_+ \sigma_-^{e}\rangle$ is determined by
full equations of motions for these quantities, which, for instance,
also include external drive of frequency $\omega_{1}$. However, it
can be shown that the simplified equations of motion (\ref{EM1}) and
(\ref{EM2}) produce correct results, while omitted terms give only
small additive contributions, which do not alter the general
conclusions on spectrum structure.

 The quantities (\ref{corr1}) and (\ref{corr2}) can be
used as inputs for Eqs. (\ref{MB3}) and (\ref{MB4}) -- they provide
additional nonclassical driving of the qubit. We also take into
account that $\omega_{01}^{e}$ can be associated with the frequency
of the drive $\omega_{2}$: $\omega_{2} \equiv \omega_{01}^{e}$. We
then switch to the rotating frame characterized by the frequency
$\omega_d$ and use the same notations as in Section II. We also
extend $t$ from $t \in (Tn,Tn+T)$ to $t\in (- \infty,+ \infty)$.
This means that $(t-Tn)$ in Eqs. (\ref{MB3}), (\ref{MB4}) must be
replaced by $\lfloor t/T \rfloor T$, where $\lfloor \ldots \rfloor$
is a floor function. The equations of motion take the form
\begin{equation}
\frac{d \langle \sigma_- \rangle}{dt} =  \langle \sigma_- \rangle
\left( -i \Delta \omega -\gamma \right)
  -\frac{i\Omega_1}{2} e^{-i \delta \omega t} \langle \sigma_z \rangle
  + \sqrt{\gamma \gamma_e} |\langle  \sigma_z \sigma_-^{e} (Tn)\rangle|
  e^{ i \delta \omega t} e^{-\left( \Gamma + \gamma_e \right) \lfloor t/T \rfloor T}, \label{MB5}
\end{equation}
\begin{equation}
\frac{d \langle \sigma_z \rangle}{dt} = -\Gamma (\langle
\sigma_z\rangle+1)  + i \Omega_1 \left(\langle \sigma_+ \rangle e^{
- i \delta \omega t} - \langle \sigma_- \rangle e^{ i \delta \omega
t}\right) + 2 \sqrt{\gamma \gamma_e} |\langle \sigma_- \sigma_+^{e}
(Tn)\rangle | \left(e^{ -2i \delta \omega t} e^{i(\delta \omega -
\Delta \omega) \lfloor t/T \rfloor T}+c.c. \right) e^{-\left( \gamma
+ \gamma_e \right) \lfloor t/T \rfloor T}.
  \label{MB6}
\end{equation}

\begin{figure}[h]\center
\includegraphics[width=.4\linewidth]{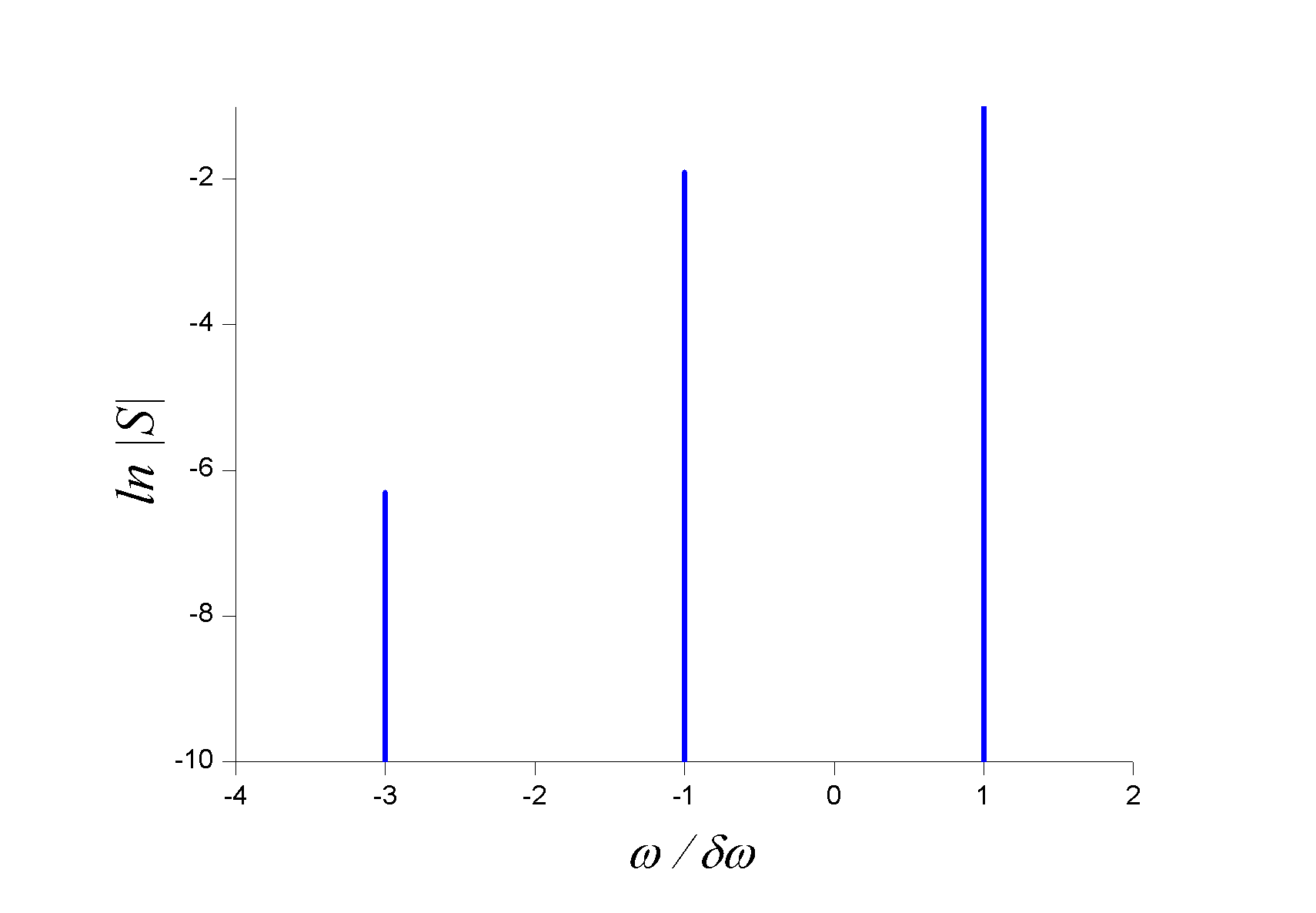}
\caption{Spectral components of $\langle \sigma_- \rangle$ in the
case  of qubit irradiation by a coherent wave together with the
superposition of Fock states with 0 and 1 photon (see in the text).}
\label{spectra3}
\end{figure}

We treat these equations as follows. Within each time interval $t
\in (Tn,Tn+T)$, functions of the form $e^{-\left( \gamma + \gamma_e
\right) \lfloor t/T \rfloor T}$ are approximated as 1 at $t-Tn
\lesssim \gamma^{-1}, \gamma_e^{-1}$ and $0$ otherwise (step-like
irradiation by the nonclassical signal). For the first interval of
time, it is readily seen from the above equations that $\langle
\sigma_- \rangle$ is a superposition of three contributions
proportional to $e^{ i \delta \omega t}$, $e^{ - i \delta \omega
t}$, and $e^{ -3 i  \delta \omega t}$, while $\langle
\sigma_z\rangle$ is a superposition of terms of the form $e^{ 2 i
\delta \omega t}$ and $e^{ - 2i \delta \omega t}$. For the second
time interval, when only a classical drive acts on the qubit,
$\langle \sigma_- \rangle$ contains only a contribution of the form
$e^{ -i \delta \omega t}$. Thus at low frequencies $\omega \ll
\gamma^{-1}$ there appear only three spectral components of $\langle
\sigma_- \rangle$. This is due to the limitation of the photon
number in the nonclassical signal this situation being totally
different from the previously considered setups.

An equivalent qualitative picture can be obtained  by considering a
stationary state solution and neglecting time derivatives in the
right-hand sides of Eqs. (\ref{MB5}) and (\ref{MB6}). The solution
can be represented as
\begin{equation}
\langle \sigma_- \rangle = c_{1}e^{ i \delta \omega t}+ c_{-1}e^{ -i \delta \omega t} + c_{-3}e^{ -3 i  \delta \omega t}.
  \label{stst}
\end{equation}
The expressions of coefficients $c_{1}$, $c_{-1}$,  and $c_{-3}$, in
general case, are rather cumbersome so we present them only for
$\Delta \omega = 0$ and in leading order in $\Omega_{1}/\Gamma$:
\begin{equation}
c_{1} \simeq - \sqrt{\frac{\gamma_e}{\gamma}}\frac{\sin \theta}{2},
  \label{coef1}
\end{equation}
\begin{equation}
c_{-1} \simeq \frac{i \Omega_{1}}{2\gamma},
  \label{coef2}
\end{equation}
\begin{equation}
c_{-3} \simeq \frac{ \Omega_{1}^2}{2\gamma^2}  \frac{\sqrt{\gamma
\gamma_e}}{\Gamma} \frac{\sin \theta}{2} \left( e^{-(\gamma +
\gamma_e)[t/T]T} e^{i\delta\omega \lfloor t/T \rfloor T}+
e^{-(\Gamma + \gamma_e)\lfloor t/T \rfloor T}\right).
  \label{coef3}
\end{equation}

\begin{figure}[h]\center
\includegraphics[width=.15\linewidth]{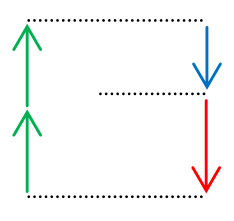}
\caption{Schematic image of a multiphoton process  resulting in a
side peak in the emission spectra at $3\delta \omega$ under the
qubit irradiation by the classical signal and quantum superpositions
of Fock states with 0 and 1 photons. Green arrows correspond to
photons from the coherent wave with energies $\omega_{1}$. Blue
arrow correspond to the single photon with energy $\omega_{2}$. Red
arrow indicates photon emission with energy $2\omega_{1} -
\omega_{2} = \omega_{d} + 3 \delta \omega$.} \label{scheme2}
\end{figure}

Thus, at low frequencies $\omega \ll \gamma^{-1}$ there  appear only
three spectral components of $\langle \sigma_- \rangle$. They are
shown in Fig. \ref{spectra3} at $\Omega_1 = 0.15 \Gamma$, $\Delta
\omega =0$, $\Gamma = 2\gamma$, $\gamma=\gamma_e$, $\nu = 1/2$,
$T=5/\gamma_e$. The only side component is given by the third term
in the right-hand side of Eq. (\ref{stst}). It appears due to the
absorption of two photons of the coherent wave with frequency
$\omega_{1}$ and the emission of a single photon at frequency
$\omega_{2}$, since there can be no more than a single photon in the
second signal which is fundamentally nonclassical. This process is
illustrated in Fig. \ref{scheme2}. The spectrum therefore is totally
different from the spectrum in the case of two coherent waves
mixing, which was described in Section II, see Fig. \ref{spectra}.
Note that $c_{-3}$ is proportional to $(\Omega_{1} / \Gamma)^2$ that
is consistent with the fact that two photons from the coherent wave
are mixed with zero or one photons of the nonclassical field within
each time "window" $T$. Another interesting observation is that the
effect of the nonclassical signal is strongest at $\nu = 1/2$ and
not at $\nu = 1$. This is due to the fact that the single-photon
Fock state contains no information about phase.

\section{Conclusions}

To conclude, we considered theoretically wave mixing between the
classical monochromatic signal and a nonclassical light. The mixing
occurs due to the interaction of two photon fields on a single qubit
that gives rise to elastic multiphoton processes. Two particular
examples of nonclassical light were addressed: broadband squeezed
light that can be produced by the degenerate parametric amplifier
and a periodically excited superposition of Fock states with 0 and 1
photons that can be generated by a single-photon source.

The spectrum for the emitted light, which contains side peaks
attributed to nonlinearities of various orders, is distinct from the
similar spectrum in the case of qubit irradiation by two classical
drives. The reason is that nonclassical photon fields are
characterized by zero occupancies of certain Fock states. For
example, in the case of a finite-bandwidth squeezed light, only
multiphoton processes involving correlated pairs of the squeezed
field contribute to side peaks amplitudes. Such a restriction is
even more strick for Fock states with 0 and 1 photons, so that only
a single side peak appears in this case.

Thus, the amplitudes of side peaks can be used to probe nonclassical
light statistics. The key idea is that light, whose statistical
properties have to be determined, must be mixed with the classical
signal on a single artificial atom. The absence of some peaks in the
elastic spectrum of the emitted light compared to the case of mixing
of two classical signals shows that the first signal is strongly
nonclassical, since occupancies of certain Fock states must be zero
for these peaks to be absent.


\section*{Acknowledgements}
We thank A. M. Satanin, A. A. Elistratov, E. S. Andrianov, O. V.
Kotov, and D. S. Shapiro for very useful discussions. W. V. P.
acknowledges a support from RFBR (project no. 19-02-00421).

\renewcommand{\appendixname}{Appendix}
\appendix

\section{Qubit irradiation by a coherent wave and quantum superposition of vacuum and one photon: equations of motion}

\subsection{Hamiltonian and preliminaries}

We consider a single qubit coupled to the waveguide, which
experiences simultaneous effect of a classical monochromatic drive
and irradiation from the source of $0 + 1$ states, which is
represented by another qubit (emitter). The equations of motion are
derived from the microscopic theory.

We assume that the one-dimensional space is discrete and the
distance between nearest points is $\delta$, while the number of
points is $N_s \rightarrow \infty$. The discreteness will be
eliminated at the end from all observables, this is a technical
issue. The creation and destruction operators $a_{R}^{\dagger }$ and
$a_{R}$ for photons in a given point $R$ are constructed from
delocalized states described by $a_{k}^{\dagger }$ and $a_{k}$ as
\begin{equation}
a_{R}^{\dagger
}=\frac{1}{\sqrt{N_s}}\sum_{k}\exp(-ikr)a_{k}^{\dagger }.
\label{ar0}
\end{equation}
Allowed $k$ take the form $- \frac{\pi}{\delta}+ \frac{2\pi m}{L}$,
where integer $m$ ranges from 0 to $N_s-1$, while $L=N_s \delta$ is
the system's length. Thus, the maximum $\omega_k$ is $\omega_{max} =
\pi c/\delta$. The difference between two closest values of energy
is $2\pi c / L$, so that the density of energy states is $\rho =
L/2\pi c$.

The Hamiltonian of the whole system can be represented as
\begin{equation}
H = H_{phot}+ H_q+ H_{int}+ H_e+ H_{int}^{(e)},\label{Htot}
\end{equation}
where
\begin{equation}
H_{phot} = \sum_k \omega_k a_{k}^{\dagger }a_{k},\label{Hphot}
\end{equation}
is a photon Hamiltonian. The second term $H_q$ is the Hamiltonian of
the qubit under the classical drive
\begin{equation}
H_q = \frac{\omega_{01}}{2} \sigma_z - f_q(t) (\sigma_+ +
\sigma_-).\label{Hq}
\end{equation}
The third term $H_{int}$ represents an interaction between the qubit
placed at $r$ and the photon field
\begin{equation}
H_{int} = \frac{1}{\sqrt{N_s}} \sum_k (e^{-ikr}g_k^{*}a_{k}^{\dagger
}\sigma_- +  e^{ikr} g_k a_{k}\sigma_+),\label{Hint1}
\end{equation}
where $r$ is qubit coordinate and $g_k$ is an interaction constant
defined as
\begin{equation}
g_k = - i \sqrt{\frac{\omega_k}{2\epsilon_0 \delta}} \mu.\label{g_k}
\end{equation}
The fourth term $H_e$ is the emitter Hamiltonian
\begin{equation}
H_e = \frac{\omega_{01}^{(e)}}{2} \sigma_z^{(e)}.\label{He}
\end{equation}
 The firth term $H_{int}^{(e)}$ describes the interaction between
the emitter positioned at $r_e$ and photon field
\begin{equation}
H_{int}^{(e)} = \frac{1}{\sqrt{N_s}} \sum_k
(e^{-ikr_e}g_k^{(e)*}a_{k}^{\dagger }\sigma_-^{(e)} +  e^{ikr_e}
g_k^{(e)} a_{k}\sigma_+^{(e)}),\label{Hint2}
\end{equation}
where $r_e$ is emitter coordinate, $g_k^{(e)}$ is defined in a
similar way as $g_k$. The term (\ref{Hint2}) is responsible for
periodical excitation of $|0 \rangle + |1 \rangle$ states through
the emitter relaxation. We do not include explicitly into
consideration an external drive which excites the emitter. Notice
that we also use a slightly nonstandard definition of the
interaction constant, since we extracted $1/\sqrt{N_s}$ from it to
the prefactor in Eqs. (\ref{Hint1}) and (\ref{Hint2}). The prefactor
is usually absorbed by $g_k$, the latter then scales as
$1/\sqrt{L}$. We stress  that finally  interaction constant will be
expressed via the qubit relaxation rate.

If the dependence of $g_k$ on $\omega_k$ can be neglected, the
interaction is determined by the local photon field $a_{r}$ at the
position of the qubit, as can be verified performing a summation in
Eq. (\ref{Hint1}):
\begin{equation}
H_{int} \sim (a_{r}^{\dagger }\sigma_- + a_{r}
\sigma_+).\label{Hint}
\end{equation}

\subsection{Equations of motion}

We are going to explore the dynamics of the system and to focus on
steady state. There is no need to introduce phenomenologically any
energy dissipation associated with the qubit within our treatment,
since dissipation is due to the decay of the qubit excited state
into continuum of photon modes. These modes as well as their
interaction with the qubit are included into the Hamiltonian.

Let us consider infinite chain of equations of motion for the qubit
in the Heisenberg picture. The equations of motion for $\langle
\sigma_- \rangle$ and $\sigma_z$ read as
\begin{equation}
\frac{d \langle \sigma_- \rangle}{dt} = -i \omega_{01} \langle
\sigma_- \rangle -i f_q \langle \sigma_z \rangle + \frac{i
}{\sqrt{N_s}} \sum_k e^{ikr}g_k\langle a_k \sigma_z \rangle,
\label{sigmaminus}
\end{equation}
\begin{equation}
\frac{d \langle \sigma_z \rangle}{dt} = 2if_q \left(\langle \sigma_+
\rangle - \langle \sigma_- \rangle \right) +\frac{2i}{\sqrt{N_s}}
\sum_k \left( e^{-ikr} g_k^{*} \langle a_{k}^{\dagger }\sigma_-
\rangle - e^{ikr} g_k \langle a_{k}\sigma_+ \rangle \right).
\label{sigmaz}
\end{equation}
They depend on higher-order correlators. The equations of motion for
them are
\begin{eqnarray}
\frac{d \langle a_k \sigma_z \rangle}{dt} = -i \omega_k \langle a_k
\sigma_z \rangle -  2 i f_{q}\left( \langle a_k \sigma_-\rangle -
\langle a_k \sigma_+\rangle \right) +  \frac{2i }{\sqrt{N_s}} \sum_p
\left(e^{-ipr} g_p^{*} \langle a_{p}^{\dagger }a_k \sigma_-\rangle -
e^{ipr} g_p \langle a_{p} a_k \sigma_+\rangle\right) + \notag \\
\frac{i}{\sqrt{N_s}} e^{-ikr} g_k^{*}\langle \sigma_- \rangle +
\frac{i}{\sqrt{N_s}} e^{-ikr_e} g_k^{e*}\langle \sigma_-^{e}
\sigma_z \rangle,
 \label{aksigmaz}
\end{eqnarray}

\begin{eqnarray}
\frac{d \langle a_{k} \sigma_+ \rangle}{dt} = i (\omega_{01}
-\omega_{k}) \langle a_{k} \sigma_+ \rangle + i f_q \langle a_k
\sigma_z \rangle - \frac{i }{\sqrt{N_s}} \sum_p e^{-ipr} g_p^{*}
\langle a_{p}^{\dagger }a_k \sigma_z \rangle -\frac{i}{2\sqrt{N_s}}
e^{-ikr} g_k^{*} \left(\langle \sigma_z \rangle + 1 \right) - \notag
\\ \frac{i }{\sqrt{N_s}}e^{-ikr_e} g_k^{e*}\langle \sigma_-^{(e)}
\sigma_+ \rangle
 \label{aksigmaplus}
\end{eqnarray}
which depend on next-order corellators. And so on.

Note that in most of the situations the dependence of $g_k$ on
$\omega_k$ can be neglected, therefore the correlators from the RHS
(right-hand side) of the above equations are reduced to the
correlators involving
 local photon field strictly at the qubit position (after the summations over $p$).

The infinite chain of equations of motion is untractable. Therefore,
certain approximations must be made.  In general, our system must be
well described by the Born-Markov approximation. It assumes that
there is no back action of the field emitted by the qubit on qubit.
We limit ourselves to the second order in $g$ that means that we
adopt Born approximation. In this case, we can truncate the infinite
chain of equations and to consider only the system
(\ref{sigmaminus})-(\ref{aksigmaplus}). We can also neglect $\langle
a_k \sigma_-\rangle$ in the right-hand side of Eq. (\ref{aksigmaz})
that is justified in the rotating-wave approximation.

Let us now concentrate on Eqs. (\ref{aksigmaz}) and
(\ref{aksigmaplus}). A simplification comes from the fact that terms
proportional to $f_q$ can be omitted in the right-hand sides of
these two equations, since they produce corrections of the order of
$\Omega_1/\Omega_{01}$. This approximation will allow us to decouple
Eqs. (\ref{aksigmaz}) and (\ref{aksigmaplus}). We also split the
correlators as $\langle a_{p}^{\dagger }a_k \sigma_-\rangle \simeq
\langle a_{p}^{\dagger }a_k \rangle \langle \sigma_-\rangle$,
$\langle a_{p}^{\dagger }a_k \sigma_z \rangle \simeq \langle
a_{p}^{\dagger }a_k \rangle \langle \sigma_z \rangle$, and $\langle
a_{p} a_k \sigma_+\rangle \simeq \langle a_{p} a_k \rangle \langle
\sigma_+\rangle$ that is justified in Born approximation. This
implies that these two quantities generated by the emitter will be
treated as inputs for the qubit's dynamics.

Now we address a couple of equations, which are Eqs.
(\ref{sigmaminus}) and (\ref{aksigmaz}). The solution of Eq.
(\ref{aksigmaz}) can be formally written in the integral form as
\begin{eqnarray}
\langle a_k (t) \sigma_z (t) \rangle = \frac{i}{\sqrt{N_s}} \int_0
^t dt' e^{i \omega_k (t'-t)}\langle \sigma_-(t')\rangle
\left(e^{-ikr} g_k^{*} + 2 \sum_p e^{-i p r}g_p^{*}\langle
a_{p}^{\dagger } (t')a_k(t') \rangle \right) + \notag \\
\frac{i}{\sqrt{N_s}} e^{-ikr_e} g_k^{e*}\int_0 ^t dt' e^{i \omega_k
(t'-t)}\langle \sigma_-^{e} (t') \sigma_z (t') \rangle.
 \label{ak2}
\end{eqnarray}
We neglected correlator of the form $a^2$ in the RHS of the above
equation, since it is irrelevant for the quantum source we here
consider provided pulses from it are well separated in time.
However, it can be relevant for overlapping pulses. We adopt now
Markov approximation which is based on the observation that there
exist a separation between fast and slow variables in the
integrands. Particularly, we insert $\langle \sigma_-(t')\rangle
\simeq \langle \sigma_-(t)\rangle e^{i \omega_{01} (t-t')}$ into the
second integral in the RHS of Eq. (\ref{ak2}).

Now we substitute Eq. (\ref{ak2}) into (\ref{sigmaminus}). The first
term in the RHS of Eq. (\ref{ak2}) provides the following
contribution to the RHS of Eq. (\ref{sigmaminus})
\begin{eqnarray}
-\frac{\langle \sigma_- \rangle}{N_s} \sum_k |g_k|^2 \int_0 ^t dt'
e^{i (\omega_k-\omega_{01}) (t'-t)}. \label{sigma3}
\end{eqnarray}
The integral in the RHS of Eq. (\ref{sigma3}), as well as other
similar integrals appearing in derivation of the equations of
motion, is evaluated as
\begin{eqnarray}
\int_0 ^t dt' e^{i (\omega_k-\omega_{01}) (t'-t)}  = \int_0 ^t dt'
\left(\cos (\omega_k-\omega_{01}) (t'-t) + i \sin
(\omega_k-\omega_{01}) (t'-t) \right) \notag \\ = \int_0 ^t dt' \cos
(\omega_k-\omega_{01}) (t'-t) + \frac{i}{\omega_k-\omega_{01}}
\left(-1 + \cos (\omega_k-\omega_{01}) t \right),  \label{integr}
\end{eqnarray}
where the last term vanishes after averaging over long time. The
first term then gives a dissipation rate, since the integral is
nonzero and equal to $\approx t$ only at $(\omega_k - \omega_{01})t
\lesssim \pi$ and the number of energy states satisfying this
condition is $\rho /t$. Therefore, the expression (\ref{sigma3}) is
reduced to $-N_s^{-1}\rho |g_{k_{01}}|^2\langle \sigma_- \rangle$.
The combination $N_s^{-1}\rho |g_{k_{01}}|^2$ is nothing but the
energy dissipation rate $\gamma$ (in absence of pure dephasing). It
is important to stress that it turns out to be independent both on
$L$ and $\delta$. The second term in the RHS of Eq. (\ref{integr})
is responsible for the Lamb shift, $\Delta_L=N_s^{-1}\sum_k |g_k|^2
(\omega_k - \omega_{01})^{-1}$. We absorb it into the definition of
$\omega_{01}$ in Eq. (\ref{sigmaminus}). Finally, the equation
(\ref{sigmaminus}) in Born-Markov approximation takes the form
\begin{eqnarray}
\frac{d \langle \sigma_- \rangle}{dt} = -i \omega_{01} \langle
\sigma_- \rangle -i f_q \langle \sigma_z \rangle  +  \sqrt{\gamma
\gamma_e}\langle \sigma_-^{e}  \sigma_z \rangle - 2\gamma \langle
\sigma_- \rangle \langle a_r^{\dagger} a_r \rangle - \gamma \langle
\sigma_- \rangle. \label{sigma31}
\end{eqnarray}

Let us consider Eqs. (\ref{sigmaz}) and (\ref{aksigmaplus}). The
solutions of Eq. (\ref{aksigmaplus}) can be formally written in the
integral form as
\begin{eqnarray}
\langle a_k (t) \sigma_+ (t) \rangle =  - \frac{i g_k^{*}e^{-ikr}
}{2\sqrt{N_s}}  \int_0 ^t dt' e^{i (\omega_k-\omega_{01})
(t'-t)}(\langle \sigma_z(t')\rangle+1) \rangle - \notag \\
\frac{i}{\sqrt{N_s}} \int_0 ^t dt' e^{i (\omega_k-\omega_{01})
(t'-t)} \sum_p g_p^{*} e^{-ipr} \langle a_{p}^{\dagger}(t')a_k(t')
\rangle- \frac{i }{\sqrt{N_s}}e^{-ikr_e} g_k^{e*} \int_0 ^t dt' e^{i
(\omega_k-\omega_{01}) (t'-t)}\langle \sigma_-^{(e)} (t') \sigma_+
(t') \rangle.
 \label{ak12}
\end{eqnarray}
In Markov approximation, $\langle \sigma_z(t')\rangle$ in the
integrand can be replaced by $\langle \sigma_z(t)\rangle$.

We substitute Eq. (\ref{ak12}) to Eq. (\ref{sigmaz}) and collect all
the terms. Within Markov approximation, we obtain
\begin{eqnarray}
\frac{d \langle \sigma_z \rangle}{dt} =  - \Gamma (1+\langle
\sigma_z \rangle) - 2 \Gamma \langle \sigma_z \rangle \langle
a_r^{\dagger} a_r \rangle + 2if_q \left(\langle \sigma_+ \rangle -
\langle \sigma_- \rangle \right)+ 2 \sqrt{\gamma
\gamma_e}\left(\langle \sigma_+^{(e)}\sigma_- \rangle + c.c.\right),
\label{sigmazxryu}
\end{eqnarray}
where $\Gamma = 2\gamma$ (pure dephasing has been neglected) .

 Thus, Eqs. (\ref{sigma31}) and
(\ref{sigmazxryu}) provide a pair of the equations of motion for the
qubit coupled to the emitter. Notice that, in our case, $\langle
a_r^{\dagger} a_r \rangle$ in RHS of both equations can be omitted.
Indeed, for the qubit under the mixed drive, these quantities are
nonzero only within each time window of duration $\sim 1/\gamma_e$
after the emitter relaxation. The mean value $\langle a_r^{\dagger}
a_r \rangle$ carries no information about the phase, so it does not
influence our qualitative result on the structure of the spectrum.
If needed, $\langle a_r^{\dagger} a_r \rangle$ can be found from the
equations of motion for the emitter.

\end{document}